\documentclass[%
 aip,
 rsi,%
 amsmath,amssymb,
 reprint,%
]{revtex4-2}

\makeatletter
\def\switch@tabular{\class@info{REVTeX tabular patching disabled (post-2023 kernel)}}
\let\switch@array\switch@tabular
\makeatother

\usepackage{graphicx}
\usepackage{booktabs}
\usepackage{array}
\usepackage{tabularx}
\usepackage{xcolor}
\usepackage[colorlinks=true,allcolors=blue]{hyperref}
\usepackage{gensymb} % \degree, \celsius
\usepackage[htt]{hyphenat} % allow long \texttt{} paths to break across lines
\newif\ifdraftnotes
\ifdefined\DRAFTNOTES\draftnotestrue\fi
\newcommand{\todo}[1]{\ifdraftnotes{\color{red}\textbf{[TODO: #1]}}\fi}

\newcolumntype{L}[1]{>{\raggedright\arraybackslash}p{#1}}
\newcolumntype{R}[1]{>{\raggedleft\arraybackslash}p{#1}}

\graphicspath{{figures/}}
\begin{document}

\title{Retrofitting a commercial RF induction generator into a
computer-controlled, vacuum and gas integrated annealing system for reactive-metal
grain growth}

\author{Sterling G. Baird}% joint corresponding author per S. Baird (PR #12,
% 2026-08-11): "should be sterling.baird@byu.edu, make both Ronnie and
% myself corresponding authors." Email corrected from the earlier
% sgbaird@byu.edu guess to the address S. Baird specified.
\email[Author to whom correspondence should be addressed: ]{sterling.baird@byu.edu}
\affiliation{Department of Mechanical Engineering, Brigham Young University,
Provo, Utah 84602, USA}
\author{Ryan Weber}
\affiliation{Department of Mechanical Engineering, Brigham Young University,
Provo, Utah 84602, USA}
\author{Christopher Nyborg}
\affiliation{Department of Mechanical Engineering, Brigham Young University,
Provo, Utah 84602, USA}
\author{Ronald Guymon}% joint corresponding author per S. Baird (PR #12,
% 2026-08-11); email rguymon2@byu.edu per R. Guymon's 2026-08-11 designation.
\email[Author to whom correspondence should be addressed: ]{rguymon2@byu.edu}
\affiliation{Department of Mechanical Engineering, Brigham Young University,
Provo, Utah 84602, USA}
\author{Gage Erickson}
\affiliation{Department of Mechanical Engineering, Brigham Young University,
Provo, Utah 84602, USA}
\author{Oliver Johnson}% last author
\affiliation{Department of Mechanical Engineering, Brigham Young University,
Provo, Utah 84602, USA}

\date{\today}

% Abstract rewritten 2026-08-11 per R. Guymon (PR #12): CAMRI framework
% (results first, impact close), under AIP's 250-word cap, plain sentences,
% SEM/DAQ definitions and the repo URL removed (URL lives in Data
% Availability). The "soak" and "sample assembly" glosses cut here moved to
% those terms' first body mentions.
\begin{abstract}
High-temperature vacuum annealing near a metal's melting point drives
controlled grain growth, but it normally requires an expensive turn-key
vacuum induction furnace. We retrofit a bare commercial radiofrequency
(RF) induction generator with computer power control through LabVIEW,
dual-wavelength optical temperature feedback, and a high-vacuum
quartz-tube chamber with inert-gas backfill. A machined graphite crucible
doubles as a susceptor, an RF-absorbing element that heats the specimen
it encloses. The only requirement is a monotonic analog power-control
input, so the retrofit transfers between generators. Our reference build
uses a 6\,kW solid-state generator. Surprisingly, nickel annealed in this
system produced high-quality electron backscatter diffraction (EBSD)
patterns with zero specimen preparation. Samples went directly from the
furnace to the electron microscope without grinding, polishing, or
etching. The EBSD maps indexed hundreds of equiaxed grains delineated by
deep thermal grooves. A modified sample assembly extends the furnace to
ceramics that do not couple to the RF field. It coarsened
yttria-stabilized zirconia (YSZ) grains in 45\,min at 2500\,\celsius,
compared with 228\,h at 1600\,\celsius{} in a conventional box furnace.
For a fixed configuration, the power--temperature calibration is linear
($R^2 = 0.991$) from 1200 to 1400\,\celsius. Eight 12\,h nickel anneals
at 1200\,\celsius{} reproduced their soak temperature to
$1201.2 \pm 1.3$\,\celsius, and 40\,h soaks at 1325\,\celsius{} remained
stable. Complete design files, a bill of materials, control software, and
data are openly available. This gives laboratories an affordable route to
near-melting-point annealing with direct anneal-to-EBSD characterization.
\end{abstract}

\keywords{induction furnace; grain growth; vacuum annealing; LabVIEW;
pyrometer; open-source hardware}

\maketitle

\ifdraftnotes
\noindent\fbox{\parbox{\linewidth}{\small\textbf{Status:} restyled
("real person") variant of paper.tex; content is identical, prose is
rewritten in the Extension-to-ceramics style. \todo{}-marked items flag
values that still need to be confirmed before submission.}}
\fi

%==============================================================
\section{Introduction}
%==============================================================

High-temperature annealing near a metal's melting point drives controlled
grain growth. This is central to studying microstructure--property
relationships in metals such as nickel and
iron~\cite{liu2021studyofgrain,mohrbacher2026applicationofmicroalloying,alogab2007theinfluenceof,tian2009experimentalandsimulation,moore2017graincoarseningbehaviour}.
Growing grains without oxidizing the sample requires temperatures of 1400 to
1500\,\celsius{} along with high vacuum or an inert
atmosphere~\cite{zhornyak1982reductionanddecarburization,muramatsu2005gascontaminationdue}.
Commercial vacuum induction furnaces that meet these requirements are
expensive, closed, turn-key systems.
% RSI policy (relayed by R. Guymon, PR #12, 2026-08-13): prices cannot be
% quoted anywhere in the text, and novelty is judged solely on technical
% grounds. The former "$50k to $200k or more" range and the "$38k" system
% total were removed; only generic in-passing cost mentions remain. Resistive tube and
box furnaces do not heat the charge directly. They heat primarily through radiation
from the furnace elements, so heat transfer into the charge is limited, and
they reach lower temperatures with slower ramps. Bare
industrial induction generators are widely available, but they ship without
vacuum, optical temperature feedback, or computer setpoint control.

This work adds those missing subsystems as an open, documented retrofit. A
bare RF induction
generator~\cite{rudnev2017handbookofinduction,lucia2014inductionheatingtechnology}
is retrofitted into a full annealing system with computer power control,
closed-loop optical temperature feedback, a high-vacuum quartz-tube chamber,
and a crucible that doubles as a susceptor. The
retrofit is deliberately generator-agnostic. Its only requirement is an analog
power-control input with a monotonic command-to-power response. The pyrometer feedback and
proportional--integral--derivative (PID) calibration absorb
any nonlinearity in that response. The generator internals are not
reproduced, so a compatible generator needs to be supplied. A graphite
crucible couples strongly to the RF field and is used for the metal
grain-growth anneals. Non-coupling ceramics such as YSZ
use a modified sample assembly (the specimen, susceptor, and supports
inside the chamber) in
which graphite or tantalum serves as the susceptor, chosen for chemical
compatibility with the specimen (Sec.~\ref{sec:ysz}).

The reproducible build documented here is anchored on a modern
6\,kW solid-state generator whose controller accepts a standard analog
power setpoint. The reference generator is a CEIA ``Power Cube'' PW3-90/50
with its matching controller, heating head, and chiller. The generator
package and the retrofit components are itemized in the bill of materials
in the supplementary material.

%==============================================================
\section{System design and description}
%==============================================================

\begin{figure*}[tb]
\centering
\begin{minipage}[b]{0.60\textwidth}
\raggedright{\textbf{(a)}}\\[1pt]
\centering
\includegraphics[width=\linewidth]{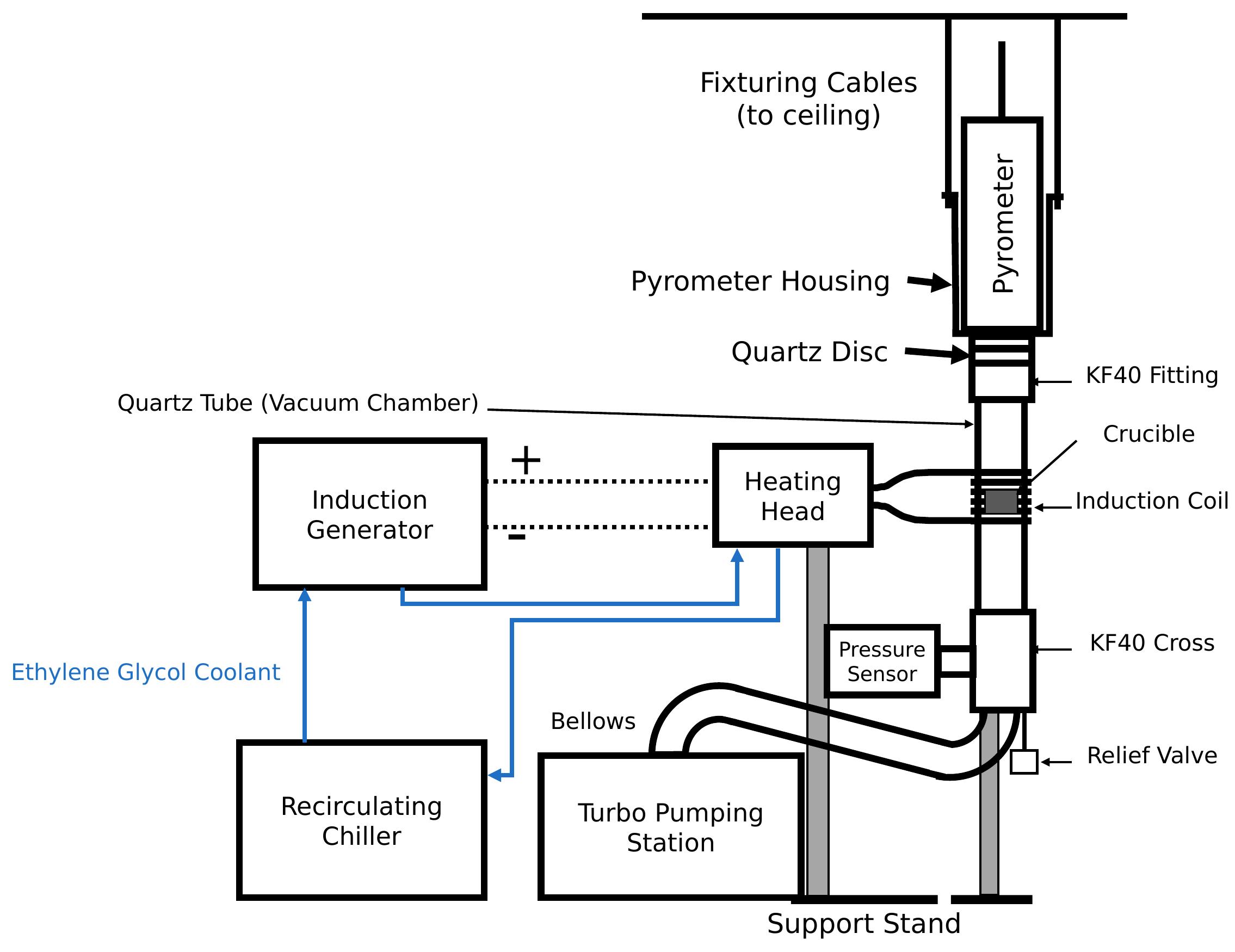}
\end{minipage}\hfill
\begin{minipage}[b]{0.365\textwidth}
\raggedright{\textbf{(b)}}\\[1pt]
\centering
\includegraphics[width=\linewidth]{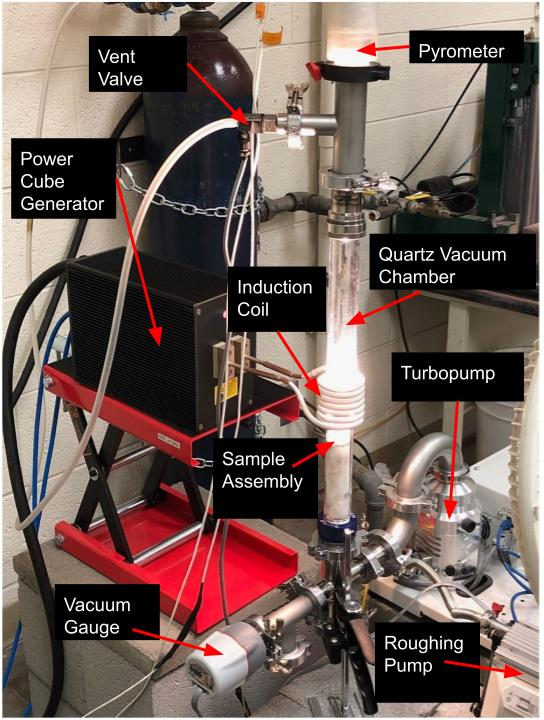}
\end{minipage}
\caption{The annealing system. (a)~System overview. The generator drives the
liquid-cooled heating head and work coil around the vertically suspended
quartz-tube chamber. The quartz tube is joined by KF40 flanges to the
ceiling-suspended pyrometer housing above and to the vacuum hardware below.
A smaller support stand under the vacuum chamber supports the vacuum
hardware from below.
An overpressure relief valve and a flexible bellows lead past the pressure
sensor to the turbopumping station. (b)~The assembled, functioning furnace
with the main components labeled. This shows the quartz chamber glowing at
the sample-assembly position. The pyrometer sights down the chamber axis, and
the vent valve admits the argon backfill. The gauge, turbopump, and
separately supported roughing pump complete the vacuum path. Relative to
closed turn-key vacuum induction furnaces, this design costs far less, is
modular (hand-clamped KF-flanged vacuum hardware and an interchangeable
sample assembly), transparent (customizable LabVIEW control software and design
files), and it reaches a high-purity annealing environment by pumping the
chamber to high vacuum and then backfilling with flowing argon.}
\label{fig:overview}
\end{figure*}

The system comprises five subsystems: heating, vacuum and gas handling,
optical temperature sensing, computer control, and mechanical support
(Fig.~\ref{fig:overview}(a)).
Figure~\ref{fig:overview}(b) shows the assembled system at operating power.

The heating subsystem is a 6\,kW solid-state induction generator driving a
liquid-cooled copper work coil ~\cite{vankan2024electromagneticinductionheating,xu2026designandoptimization}. In our case, we procured a coil from CEIA that is approximately 3\,in tall with a
2.5\,in inner diameter and 6.5
turns, tailored to our vacuum
chamber geometry.
The coil conductors carry both the RF current and the coolant. Ethylene
glycol coolant flows in series from the chiller through the generator and
then through the heating head and work coil before returning to the chiller.
The vacuum and gas subsystem starts from the 35\,mm inner-diameter
fused-quartz tube that forms the chamber. The tube connects through a KF40
flange (a standard clamp-sealed vacuum flange) and a flexible bellows to a
compact turbopumping station that
reaches 10$^{-6}$ to 10$^{-8}$\,Torr. An automated, electrically triggered
purge vent valve fed from an inert-gas regulator allows controlled venting and
backfilling.
These components support a range of operating atmospheres. The furnace can
run at low or high vacuum, and the chamber can instead be backfilled with
inert gas and continuously supplied either at a precisely controlled rate,
for example 20 standard cubic centimeters per minute (SCCM), or as a large
purge flow. The turbopump can be left running during backfilled operation or
stopped.
The atmosphere matters for
graphite-crucible anneals~\cite{kanda2007developmentofhightemperature,inaba1986effectsofthe}.
% Citations moved here from the vent-valve hardware sentence (R. Guymon,
% PR #12, 2026-07-24): Kanda 2007 documents a low-oxygen
% atmosphere-controlled furnace for high-temperature processing, and
% Inaba 1986 shows annealing-atmosphere conditions govern oxidation of a
% Ni alloy -- both support the atmosphere-matters claim, not the valve
% hardware they were previously attached to.
The best results were
achieved by pumping to high vacuum and then backfilling with argon through
the purge vent valve. The chamber was purged first, and a constant argon flow of
20\,SCCM was then maintained for the
duration of the anneal using a mass flow controller. Many of the reported nickel grain-growth anneals
ran under this continuous flow.
% "all ran" -> "Many of the reported ... ran" per R. Guymon's edited PDF
% (PR #12, 2026-07-23/24) -- resolves the contradiction with Table S1's
% no-flow calibration runs (AI-edit review finding No. 1).
Heating under vacuum alone tended to sublime graphite onto the chamber
walls. The highest-temperature ceramic work (Sec.~\ref{sec:ysz}) used the
same continuous argon supply, with the chamber evacuated and then filled
with argon while the turbopump was running. This helped avoid oxidation
during heating.

Temperature is sensed optically. A dual-wavelength ratio
pyrometer~\cite{tapetado2016twocolorpyrometerfor,belikov2024fastmultiwavelengthpyrometer}
with an 800 to 2500\,\celsius{} sensing range views the sample through a
quartz optical window. This provides a non-contact signal that is robust to
emissivity changes and to the RF field that would corrupt
thermocouples~\cite{ham2016insituspectralemissivity,suleiman2025improvingpyrometryof,usamentiaga2014infraredthermographyfor}.
Control is implemented in LabVIEW, with virtual instruments (VIs) for
manual control, automated
ramping and PID tuning. The generator's power setpoint can be
commanded either by a 0--5\,V voltage signal or by a 4--20\,mA current-loop
signal. Which input is available depends on the generator model and its
configuration.
% Scoped per R. Guymon's question (PR #12, 2026-07-23): "typically
% customizable" was an unverifiable blanket claim about manufacturers
% (traceable to an earlier review comment); this wording keeps the
% either/or point without it.
In
this build, the data-acquisition (DAQ) device's 0--5\,V analog output
drives a voltage-to-current loop conditioner, and the conditioner's
4--20\,mA output drives the generator's power-setpoint input. The
pyrometer's analog output returns to a DAQ input, which closes the PID loop.
Below the pyrometer's 800\,\celsius{} detection floor, the VI
ramps open-loop. Once the pyrometer returns a sustained valid reading, the
VI switches to closed-loop PID control based on the optically-measured
temperature~\cite{meng2023designofvacuum,rapoport2006optimalcontrolof}.
Mechanically, a bolted support stand carries the generator's heating head
and centers the work coil on the crucible position. The quartz tube and the vacuum hardware below
it are joined by KF40 flanges to the pyrometer housing, which is suspended
from the ceiling by cables. A second, smaller support stand sits under the
vacuum chamber and supports the vacuum hardware from below.
% Support stand under the vacuum chamber added per R. Guymon (PR #12,
% 2026-07-23/24), reversing the earlier "no support stand" statement; drawn
% in Fig. 1(a) via paper/add_chamber_stand_schematic.py. The load split
% between the ceiling cables and this stand is not specified -- confirm
% with C. Nyborg if it matters for the text.
The sample assembly is not attached to any of these flanges. It
simply sits inside the quartz tube, resting on a retaining ring of a KF40
flange.

Metal charges in the graphite crucible reach the nickel and iron
grain-growth range of 1400 to 1500\,\celsius, with demonstrated soaks
(constant-temperature holds) from minutes to 40\,h. The modified ceramic sample assembly extends the working
range to 2500\,\celsius. When the retrofit is moved to a different
generator, a user would need to re-determine optimal parameters specific to their system; however, the general framework is agnostic to the generator type and brand.

\subsection{Graphite crucible / susceptor}

\begin{figure}[tb]
\centering
\includegraphics[width=0.92\linewidth]{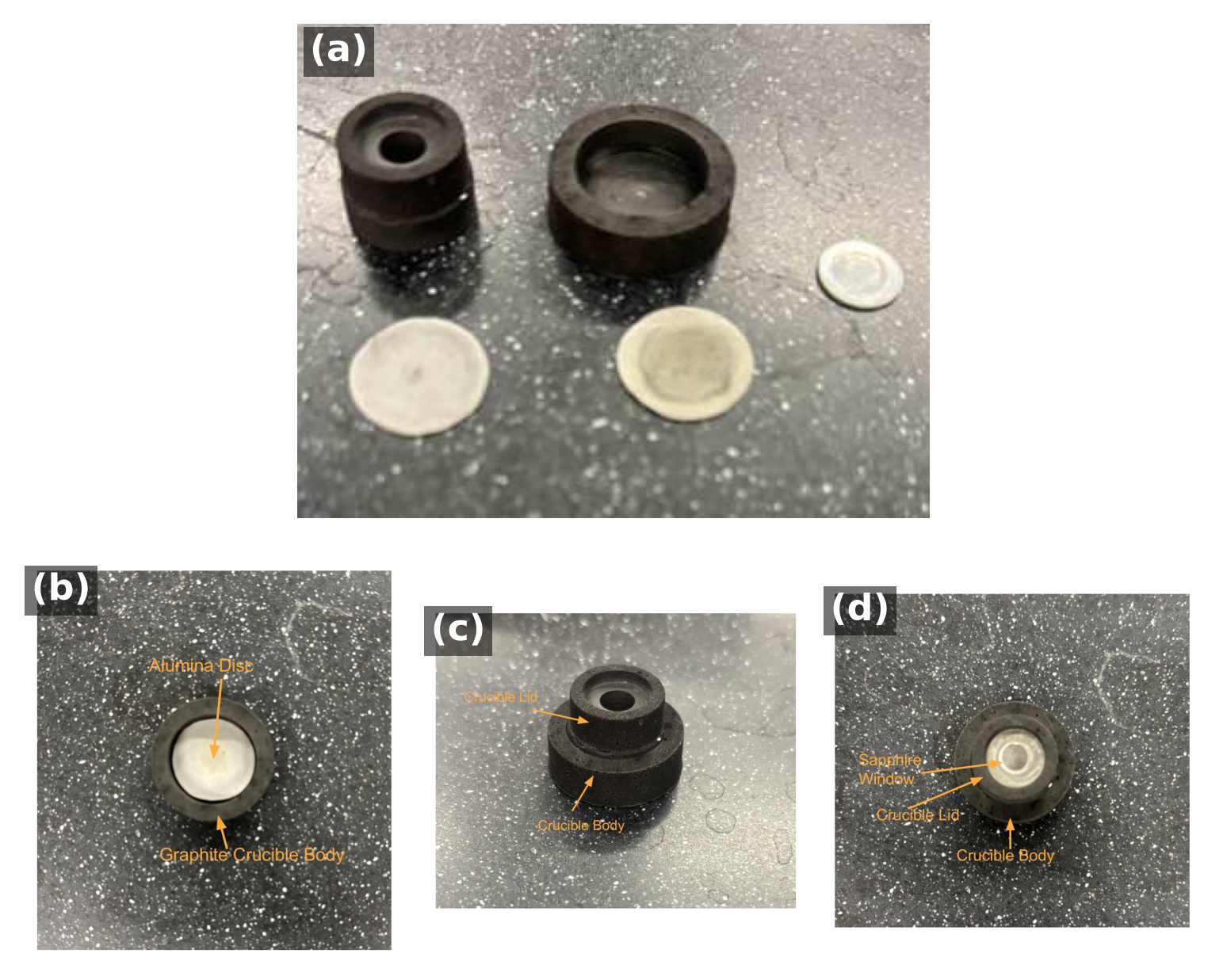}
\caption{The two-piece graphite crucible/susceptor used for the nickel
anneals reported in this work. (a)~Fully disassembled. Top left is the
crucible lid, top middle is the bottom of the crucible that acts as the
specimen carrier, far right is the sapphire window, bottom left and middle
are the alumina discs. (b--d)~Specimen-loading sequence.
(b)~The specimen is placed between two alumina discs in the crucible body.
(c)~The graphite lid is seated on top. (d)~The sapphire window closes the lid
bore, leaving an optical path for the pyrometer.}
\label{fig:crucible}
\end{figure}

Metal specimens are held in a two-piece graphite crucible machined in-house
from 3000\,\celsius-grade graphite stock. It consists of a cup body and a
lid with a central bore, which contains the charge while leaving an
optical path for the pyrometer. Figure~\ref{fig:crucible}(a) shows the
disassembled parts. The graphite couples strongly to the
induction field and acts as a susceptor, heating the contained metal
specimen by conduction and radiation. The specimen is sandwiched between two
alumina discs in the cup, the lid sits on top, and a sapphire window closes
the lid bore (Fig.~\ref{fig:crucible}(b--d)). Sapphire is used for the
window to increase transmittance in the wavelength ranges emitted from the
sample relative to the pyrometer and because it has a higher melting point than quartz. The measured dimensions of every piece are
given in the
supplementary material. Inside the vacuum chamber the crucible rests on an
alumina support tube that positions it at coil height. The tube has holes
bored laterally along its outer perimeter to aid evacuation, and its lower
end rests on top of a Teflon tube at the KF40 fitting below.
% Support-path description per S. Baird (PR #12, 2026-07-22): the alumina
% support is a tube (not a rod), with lateral vent holes. Corrected per
% R. Guymon (PR #12, 2026-07-23): the teflon tube is UNDERNEATH the alumina
% tube -- the alumina rests on top of it, not inside it.
% A parallel same-day wording pass states
% the sample assembly rests on a retaining ring of a KF40 flange (Sec. II);
% both statements are kept -- confirm the exact load path (does the teflon
% tube sit on that retaining ring?) with C. Nyborg.
The crucible is used primarily
for the metal grain-growth anneals. Graphite also works as a susceptor for some
ceramic charges when the contact chemistry permits.
Sec.~\ref{sec:ysz} describes how a compatible susceptor is selected for
refractory ceramics.

%==============================================================
\section{Construction and operation}
%==============================================================

Following the heating path, the generator's heating head mounts on an
adjustable platform on the bolted support stand so the coil centers on the
crucible position of the vertically suspended quartz tube. The vacuum path
begins at the both-ends-open quartz tube. Its lower end joins
the vacuum hardware below through a KF40 flange. Its upper end mates to the
pyrometer housing, a custom 3D-printed part that terminates in a KF40-sized
fitting. At that joint the chamber is sealed not by the printed part but by
a 55\,mm quartz optical disc pressed against an O-ring, and the pyrometer
sights through this disc down the chamber axis. Every flanged joint compresses an elastomer
O-ring carried on a centering ring, so the vacuum hardware assembles with
hand-tightened clamps and no tools. These seals support the 10$^{-6}$ to
10$^{-8}$\,Torr operating vacuum. From the bottom of the vacuum column, a tee
carries a 0.5\,psi overpressure relief valve that protects the quartz tube
during backfilling. A flexible bellows continues through the wide-range
vacuum gauge to the turbopumping station. The roughing pump sits on a
separate support, mechanically decoupled so its vibration does not reach the
chamber or the pyrometer sight line. Photographs of the pumping station as
installed, together with the relief valve at the base of the vacuum column,
are given in the supplementary material. On the gas path, the argon regulator
feeds the automated, electrically triggered vent valve. This is a
specialized commercial component that admits gas in a controlled way rather
than dumping atmospheric pressure into the chamber, which could damage the
spinning turbopump. A mass flow controller between them is helpful
when continuous inert gas flow is required. On the control path, electrical
isolation is maintained between the DAQ wiring and the generator interface.
Manual control is verified before automated ramping and soaking are enabled.

% "A run begins with the chamber vented..." sentence removed per R. Guymon
% (PR #12, 2026-07-23/24): the chamber is not held under vacuum between
% samples, and the turbopump-protection detail survives at the end of this
% paragraph.
A run begins with the specimen being loaded into the crucible
(Fig.~\ref{fig:crucible}(b--d)), the sample assembly lowered into the
quartz tube, and the chamber pumped down. The LabVIEW profile ramps open-loop below the pyrometer
floor, switches to closed-loop PID, and runs the programmed soak.
Afterwards the sample cools under vacuum, and the chamber is vented only
once the turbopump has stopped.

%==============================================================
\section{Performance validation}
%==============================================================

The furnace has completed more than one hundred logged anneals of Ni200
(the standard commercially pure wrought nickel grade) and
Ni (4N5, i.e., 99.995\,\% purity) nickel charges at 900 to
1400\,\celsius, with soaks from minutes to 40\,h. The supplementary material
provides links to raw run logs for specific samples.
% "thermal" scopes the claim to Sec. IV: every Sec. IV cohort has a linked
% run log in SI Table S1, but two Sec. V specimens do not (Ni_003b1a has no
% row; the Ni4N5_069 log is not in the parsed set).

Four steady-state anneals spanning 1200 to 1400\,\celsius{} were run with
one fixed configuration of Ni (4N5) specimen, quartz tube, graphite
crucible, and optics. Across these anneals, the soak-mean pyrometer
temperature follows the soak-mean power command, expressed in volts on
the DAQ's 0--5\,V output scale, monotonically and
approximately linearly. The relation is well described by
$T = 931 + 632\,V_{\mathrm{P}}$, with $T$ in \celsius{} and
$V_{\mathrm{P}}$ the power command in volts (equivalently 9 to 15\,\% of
full scale), and $R^2 = 0.991$, i.e., a high linear response (Fig.~S4 of the supplementary material). Its coefficients shift with the charge, crucible, and optical
configuration, so any modification to the system calls for recalibration.

A
Ni (4N5) anneal at a nominal 1300\,\celsius{} and 12\,h condition serves as
the representative closed-loop run. The full annealing curve, including the
ramp to temperature, is specified by the user in the control software in
advance. The recorded trace is given in Fig.~S5 of the supplementary
material. The soak held $1302.1 \pm 3.0$\,\celsius{} (std. dev.) over the 12\,h
plateau.

Eight Ni (4N5) anneals at a nominal 1200\,\celsius{} and 12\,h condition were
run under the same gas-flow configuration. They reproduced their soak
temperature to $1201.2 \pm 1.3$\,\celsius, a coefficient of variation of
0.11\,\% (Fig.~S6 of the supplementary material). Long-duration stability
was demonstrated with two Ni200 anneals at
1325\,\celsius. The first held for 20\,h at $1326.5 \pm 1.5$\,\celsius,
and the second held for 40\,h. Both traces are shown in Fig.~S7 of the supplementary material.
% SI figure numbers (S4 calibration, S5 representative, S6 repeatability,
% S7 long soak) are hard-coded here; keep in sync with SI.tex figure order.
Reported temperatures are specific to our setup. Slight variations to the
optical pathway or other conditions can affect the temperature readings. By
extension, absolute specimen temperature still depends on emissivity and
surface
condition~\cite{ham2016insituspectralemissivity,giulietti2025spectralemissivitymeasurement,kieruj2016determinationofemissivity}.

%==============================================================
\section{Microstructural validation}
%==============================================================

\begin{figure}[tb]
\centering
\includegraphics[width=0.36\linewidth]{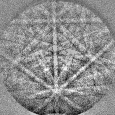}
\caption{Raw EBSD Kikuchi pattern from a furnace-annealed nickel specimen
with no sample preparation (grinding, polishing, etching, etc.). The specimen was pulled directly from the furnace chamber and
placed directly into the EBSD chamber. The pattern is reproduced exactly as
saved by the EBSD detector during orientation mapping (the detector's
$8\times8$-binned output). Even so, the crisply resolved bands and zone axes
index directly to the nickel lattice.}
\label{fig:kikuchi}
\end{figure}

The furnace exists to change microstructure, so the decisive test is what
the anneals do to real specimens. A striking feature of that test is that it
required no metallographic sample preparation at all. The EBSD specimens
shown here were never ground, polished, or etched. Samples were able to be pulled from the furnace chamber and placed
directly into the EBSD chamber with no processing (e.g., polishing), and each still returned high-quality
diffraction patterns rivaling that of high-quality polishing procedures and without scratch artifacts.
Figure~\ref{fig:kikuchi} shows such a high-quality raw EBSD pattern, also
called a Kikuchi
pattern, from an unprepared furnace-annealed nickel specimen,
reproduced exactly as the EBSD detector saved it during orientation mapping.
Even at the coarse $8\times8$ binning used for mapping, the bands and zone
axes are crisply resolved. We surmise that preparation-free EBSD on these nickel specimens works here for two reasons.
EBSD diffracts from only the top tens of nanometers of the surface, and the
long near-solidus anneal removes the deformed near-surface layer that
polishing normally exists to
remove~\cite{goldstein2018characterizingcrystallinematerials}.

\begin{figure}[tb]
\centering
\includegraphics[width=0.94\linewidth]{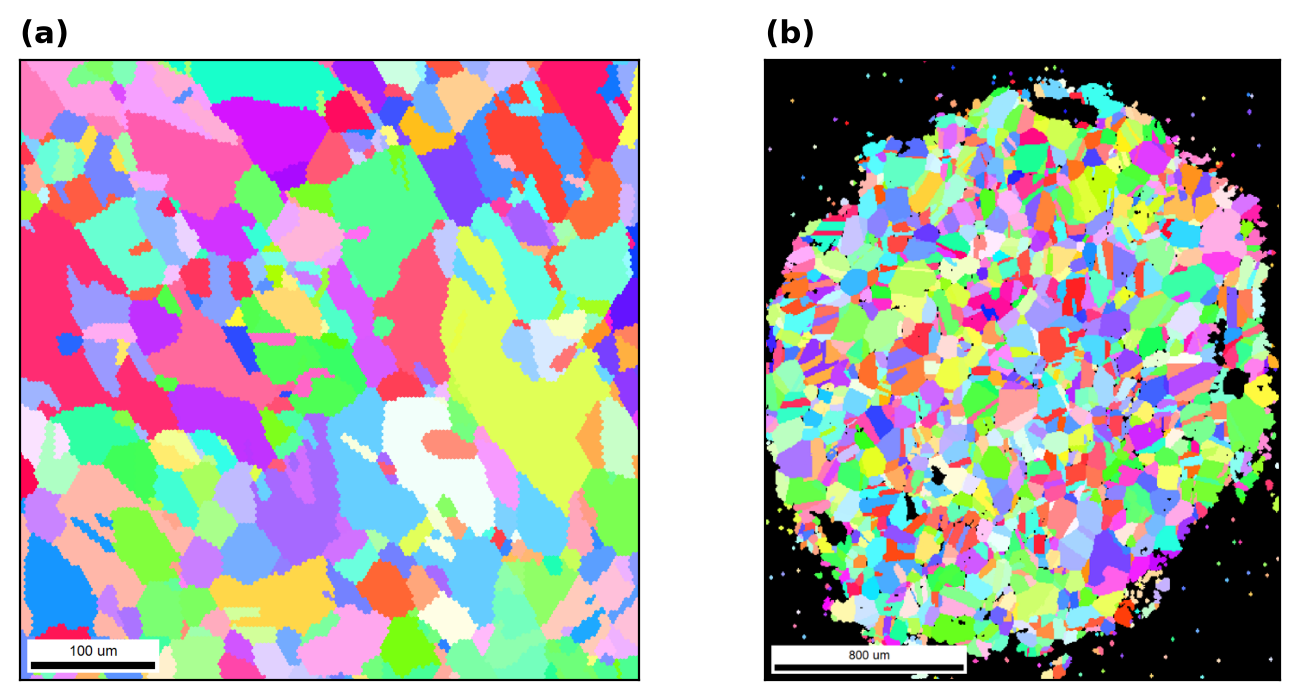}
% Panel anneal conditions are deliberately omitted here: the (a) specimen's
% run is linked in SI Table S1, but the (b) specimen (Ni4N5_069) has no
% run-log linkage in the parsed set, and the recipe should appear for both
% panels or neither (S. Baird, PR #12, 2026-07-22).
% TODO (confirm with S. Baird): at which step the (b) specimen was masked,
% so the caption's masking wording can be made more specific.
\caption{EBSD inverse-pole-figure (IPF) orientation maps of two
furnace-annealed Ni (4N5, 99.995\,\% purity) specimens, rendered from the
recorded EBSD scans. Neither specimen received any sample preparation.
(a)~A furnace-annealed specimen (100\,\textmu m scale bar). (b)~A second
furnace-annealed specimen (800\,\textmu m scale bar). This second specimen was
masked for a downstream application related to hydrogen diffusivity measurements, which is why indexed points fill only the circular unmasked region
against the black surroundings. The maps resolve hundreds of grains for
grain-size and texture analysis. Both specimens appear in the
specimen--thermal-history linkage table in the supplementary material.}
\label{fig:ebsd}
\end{figure}

\begin{figure}[tb]
\centering
\includegraphics[width=0.92\linewidth]{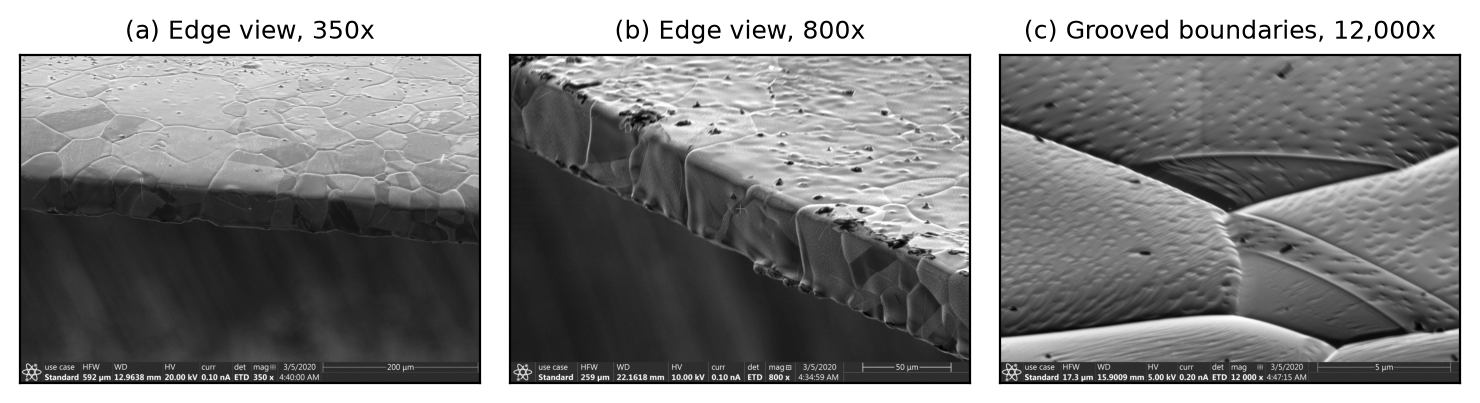}
\caption{Grain-boundary thermal grooving on a Ni (4N5) specimen after a
1200\,\celsius{} / 12\,h anneal, imaged edge-on in the as-annealed condition
after only an ultrasonic ethanol clean. The grooves have deepened enough
that they delineate the grain structure on their own. (a)~Edge view at
350$\times$ (200\,\textmu m scale bar). Several grain boundaries traverse
the full sheet thickness. (b)~Edge and top surface at 800$\times$
(50\,\textmu m scale bar). The grooves outline the surface grain structure.
(c)~Grooved boundaries at 12{,}000$\times$ (5\,\textmu m scale bar).
Multiple boundaries are grooved, and grooving is visible at several triple
junctions. The trenches deepen as grain-boundary energy equilibrates against
surface energy at the soak temperature.}
\label{fig:grooving}
\end{figure}

In addition,
the evacuated, argon-flowing, graphite-enclosed environment holds the oxygen
potential low, which plausibly keeps the surface free of an EBSD-obscuring
nickel oxide scale~\cite{wolf2021thermodynamicassessmentof}. Rapid nickel
surface diffusion at these temperatures can further smooth the
surface~\cite{maiya1967surfaceselfdiffusion,thompson2012solidstatedewettingof}.
The same surface transport produces the deep boundary grooves described
below, which can locally reduce indexing right at the
boundaries~\cite{gladstone2001grainboundarymisorientation}.
% Mechanism/precedent support: Edison Scientific literature task
% e646f435-cd6f-4489-b67d-5ca6b5a0c19b (paper/ebsd_noprep_query/
% edison_noprep_ebsd_report.md), whose suggested wording and caveats
% ("free of an EBSD-obscuring oxide scale, not necessarily oxide-free")
% these sentences follow. All five added DOIs verified against Crossref.
Inverse-pole-figure maps built
from such patterns resolve hundreds of indexed grains per specimen
(Fig.~\ref{fig:ebsd}). Each indexed color in the map is a crystallographic
orientation, so a single map records both the grain sizes and the texture of
the annealed specimen. Thermal
grooving is itself a record of the anneal. Figure~\ref{fig:grooving} shows
edge-on views of a Ni (4N5) specimen annealed at 1200\,\celsius{} for 12\,h and
cleaned only ultrasonically. The grooves have deepened enough that they
delineate the grain structure on their own, with no etching. Several grain
boundaries traverse the full sheet thickness.

\begin{figure}[tb]
\centering
\includegraphics[width=0.8\linewidth]{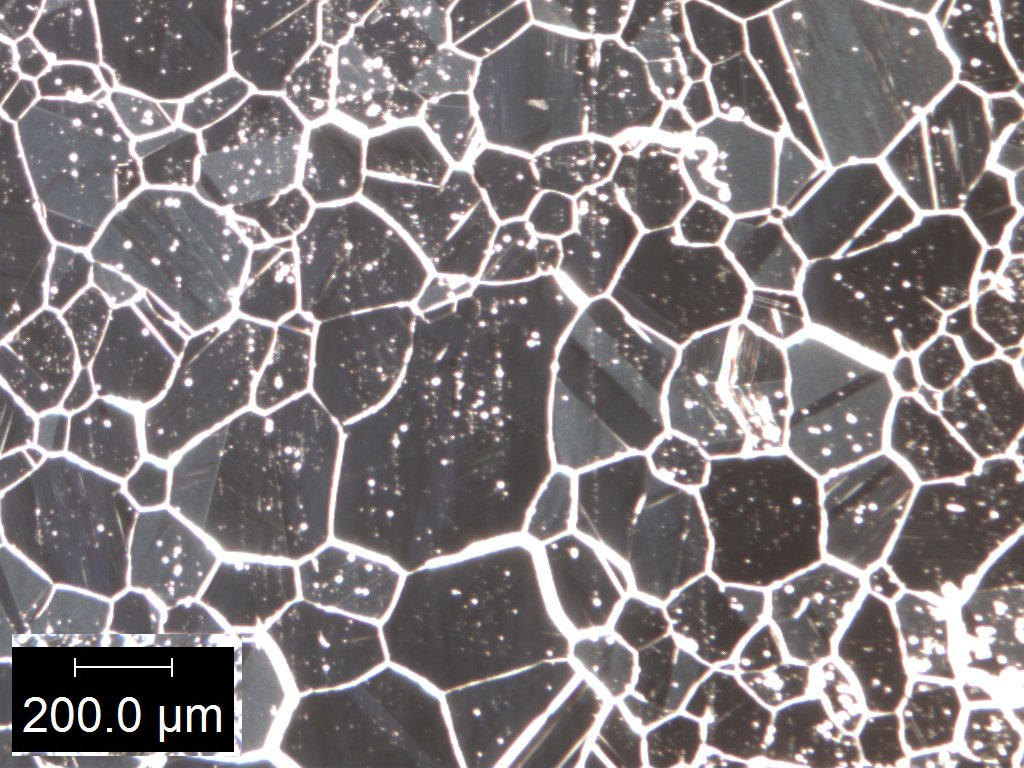}
% Image regenerated per S. Baird (PR #12, 2026-07-22): the SEM detail panel
% (b) is removed, and the optical map is cropped to the bottom-right quadrant
% of the source micrograph to exclude the overlapping yellow grain-size
% measurement text. The microscope's 200 um scale-bar box is carried over
% onto the crop at native pixel scale (no rescaling), so calibration holds
% (paper/build_microstructure_figure.py).
\caption{Optical grain map of a Ni (4N5) specimen after a
1300\,\celsius{} / 20\,h anneal undergoing a constant 20\,SCCM argon flow (200\,\textmu m scale bar), preceded by an evacuation of the chamber to high vacuum and a backfilling purge directly from high vacuum with argon. The equiaxed grain network is delineated by
thermal grooving alone.}
\label{fig:microstructure}
\end{figure}

Figure~\ref{fig:microstructure} shows the microstructure such an anneal
produces for a Ni (4N5) specimen soaked 20\,h at 1300\,\celsius{} under the
constant 20\,SCCM argon flow (preceded by an evacuation of the chamber to high vacuum and a backfilling purge), with a measured soak of approximately
1301\,\celsius.
The grains are equiaxed, and thermal grooving allows for the full grain
boundary network to be observed under an optical microscope.

\subsection{Extension to ceramic charges: YSZ}
\label{sec:ysz}

The same induction furnace can also be used to process refractory ceramics
by modifying only the materials directly in contact with the specimen (the sample assembly) while
leaving the generator, vacuum system, pyrometer, cooling system, and control
software unchanged. Materials that do not couple directly to the RF field
require a susceptor.
At the temperatures required for grains to coarsen on refractory ceramics,
the primary consideration in selecting a susceptor is its chemical
compatibility with the specimen and surrounding support materials rather
than its RF heating performance. Graphite and tantalum both serve
effectively as susceptors, with the preferred choice depending on the
operating temperature and the materials in contact with the susceptor.

\begin{figure}[tb]
\centering
\includegraphics[width=0.66\linewidth]{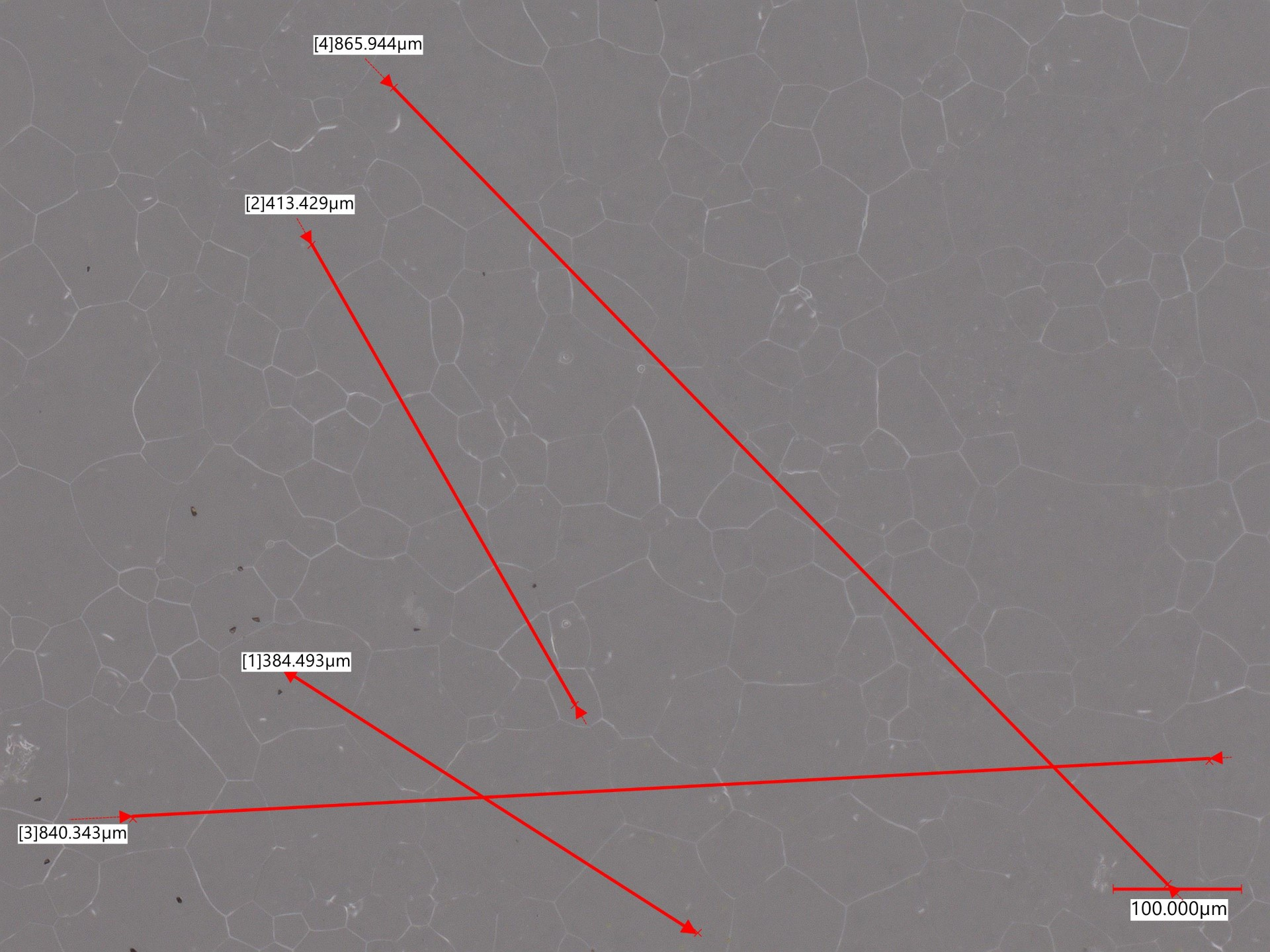}

\caption{As-recorded optical micrograph of a YSZ specimen after the
2500\,\celsius{} / 45\,min tantalum-susceptor anneal, which coarsened the
grains from approximately 20 to 90\,\textmu m. The red traces are the
microscope software's multi-grain size measurements (100\,\textmu m scale
bar).}
\label{fig:graingrowth}
\end{figure}

Grain growth in refractory ceramics is thermally activated. The rate
constant of the grain-growth law follows an Arrhenius temperature
dependence, so grain coarsening is highly sensitive to annealing
temperature~\cite{tekeli2005colloidalprocessingsintering,matsui2010phasetransformationandgrain}.
For the 8\,mol\% YSZ
specimens investigated here, grain growth from approximately 10\,\textmu m
to 80\,\textmu m required over 228\,h at 1600\,\celsius{} in a conventional
box furnace, whereas the induction furnace coarsened grains from
approximately 20\,\textmu m to 90\,\textmu m in only 45\,min at
2500\,\celsius{}
(Fig.~\ref{fig:graingrowth}).

\begin{figure}[tb]
\centering
\includegraphics[width=0.7\linewidth]{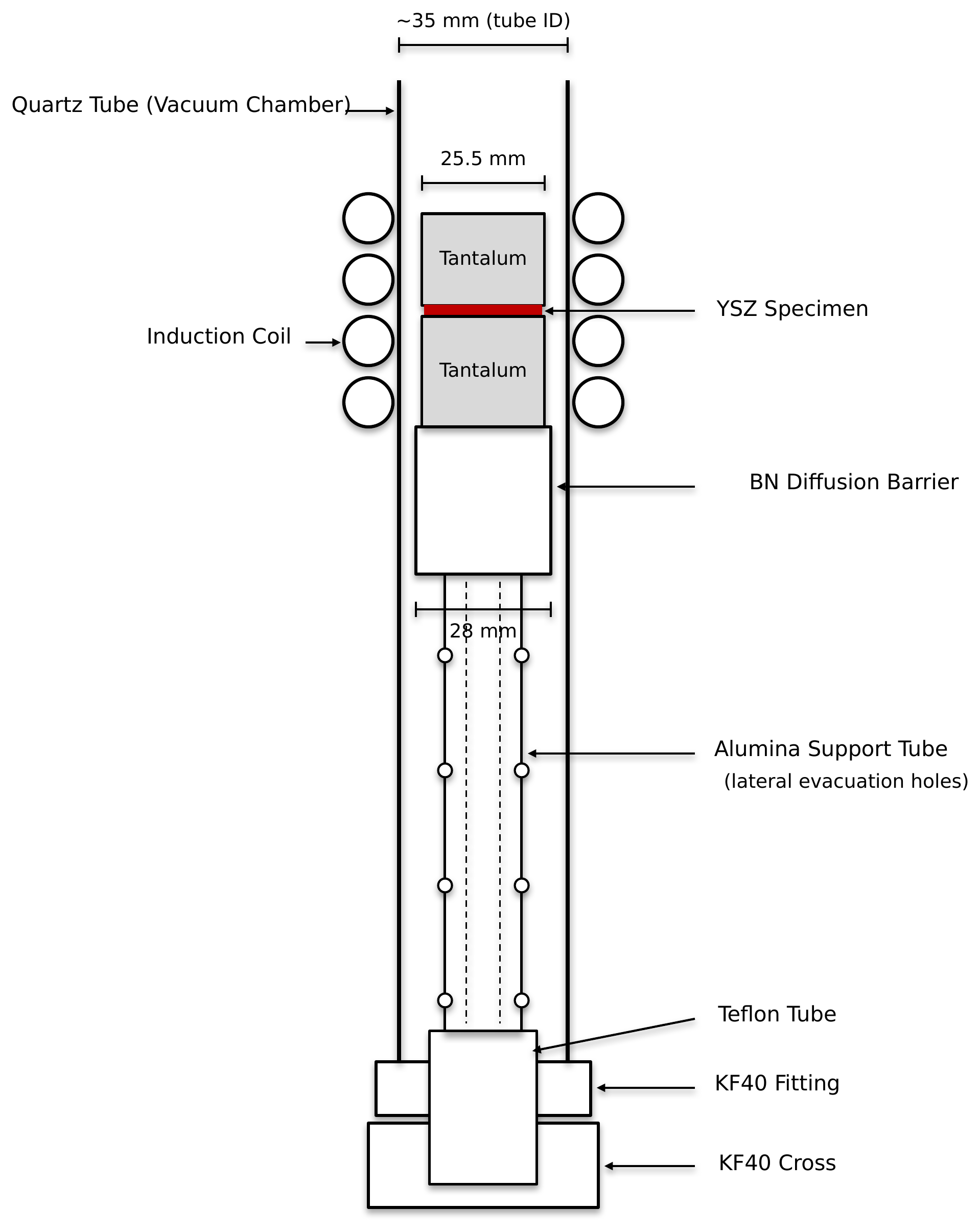}
\caption{Sample-assembly configuration for the high-temperature ceramic
anneals. A YSZ specimen is sandwiched between two 25.5\,mm tantalum
susceptor blocks inside the quartz tube. The blocks rest on a boron nitride
stub that acts as a diffusion barrier. It prevents reaction between the
tantalum and the alumina support tube below, which raises the sample
assembly to coil height. The alumina tube has holes bored laterally along
its outer perimeter to aid evacuation, and its lower end rests on top of a
Teflon tube at the KF40 fitting. The coil, vacuum, pyrometer, and control paths are
unchanged from the metal configuration.}
\label{fig:ysz-stack}
\end{figure}

Reaching this temperature range required only
minor modifications to the sample assembly
(Fig.~\ref{fig:ysz-stack}). Direct contact between graphite and alumina was eliminated because alumina is not inert
to carbon at these temperatures. Under vacuum or low oxygen potential,
carbon carbothermally reduces alumina at the interface. This evolves
carbon monoxide and forms aluminum carbides and oxycarbides that degrade
the
interface~\cite{cox1963aninvestigationof,kruesi2011solaraluminumproduction,walker2013experimentalverificationof}. The contact was eliminated by introducing a boron nitride
diffusion barrier between the two materials. Alternatively, replacing the
graphite with tantalum avoids this compatibility issue while still
providing efficient RF coupling.

%==============================================================
\section{Conclusions}
%==============================================================

A bare commercial RF induction generator can be converted into a
computer-controlled, vacuum and gas integrated, pyrometer-feedback annealing system
with a documented retrofit. The resulting system is modular, open, and
customizable. The vacuum hardware assembles with hand-tightened clamps, the
sample assembly is interchangeable, and the LabVIEW control software and
design files are provided. Because the retrofit requires only a monotonic
analog power-control input, the design transfers across generators and is
directly reusable at a small fraction of turn-key cost.

The full system and workflow deliver a high-purity annealing environment.
The strongest evidence is that, surprisingly, the annealed nickel specimens needed no
metallographic preparation at all. Specimens were annealed, removed from
the furnace, and placed directly into the scanning electron microscope,
and they returned
high-quality EBSD patterns and orientation maps with no grinding,
polishing, or etching.

Complete annealing workflows were demonstrated on both nickel and YSZ. On
a modern solid-state generator, the nickel campaign produced a
power--temperature calibration with $R^2 = 0.991$, soak temperatures
reproducible to 0.11\,\%, stable closed-loop soaks up to 40\,h, and the
controlled grain growth the system was built to produce. The YSZ campaign
highlighted the challenges of material compatibility at extreme
temperature. Every failure mode was a contact reaction among the charge,
the susceptor, and the support materials, and the interchangeable sample
assembly lets a user iterate on those material choices run by run.

High-temperature grain growth was achieved for YSZ in under an hour at
2500\,\celsius{} in this system, whereas comparable coarsening required
228\,h in a standard 1600\,\celsius{} box furnace. The system is also
relatively low-cost compared with commercial systems that reach
2500\,\celsius. In the annealed nickel, many grains and their boundaries span
the full sheet thickness, and the deep thermal grooves that mark those
boundaries point to columnar grain-growth studies as a natural next
application~\cite{zhang2007dynamicsandmechanism}.

%==============================================================
\section*{Supplementary Material}
%==============================================================
% AIP: "Please create a 'Supplementary Material' section in your paper after
% the Conclusions" (template/rsi/aip-author-instructions-2026-07-02.txt).

The supplementary material is organized in seven sections.
% SI section numbers are hard-coded here; keep in sync with SI.tex order.
Sec.~SI (hardware) gives the measured crucible part dimensions, the
work-coil drawing, and photographs of the vacuum and gas-handling
hardware. Sec.~SII (thermal performance) collects the power--temperature
calibration and the representative-soak, repeatability, and long-soak
traces. Sec.~SIII (microstructure) presents the raw Kikuchi-pattern
survey. Sec.~SIV covers the high-temperature extension, with the
tantalum-susceptor heat curves and the YSZ micrographs. Sec.~SV holds the
specimen--thermal-history linkage table, with links to raw run logs.
Sec.~SVI is the itemized bill of materials, and Sec.~SVII is the
design-file inventory.

\begin{acknowledgments}
The authors thank Kevin Cole for his support of the original
retrofit and acknowledge the Johnson group and the Brigham Young University
Department of Mechanical Engineering, which funded the capital equipment
cost. This work was supported by the National Science Foundation under
Grant No.~1610077. The authors acknowledge the BYU Electron Microscopy
Facility for providing access to the equipment and expertise that allowed
this project to be performed. This work has also been made possible, in
part, by a grant from the Simmons Research Endowment at Brigham Young
University.
% NSF grant sentence supplied by S. Baird (PR #12, 2026-07-22).
% EM Facility + Simmons Research Endowment sentences transcribed from
% R. Guymon's edited PDF (PR #12, 2026-07-23/24); R. Guymon to double-check
% the facility's prescribed standard wording.
\end{acknowledgments}

\section*{Author Declarations}

\subsection*{Conflict of Interest}

The authors have no conflicts to disclose. \todo{confirm with all authors.}

\subsection*{Ethics Approval}

% Kept per the AIP author instructions (template/rsi/aip-author-instructions
% -2026-07-02.txt): the required manuscript order includes an "author
% declarations section (conflict of interest, ethics approval, and author
% contributions)". A detailed statement is mandated only for human/animal
% research; for not-applicable work, AIP's prescribed wording (stated
% explicitly for its APB/BPR journals) is "Ethics approval not required."
Ethics approval not required. This work did not involve human or animal
subjects.

\subsection*{Author Contributions}

\todo{confirm CRediT roles} --- provisional: \textbf{Sterling G. Baird:}
Conceptualization, Software, Investigation, Writing -- original draft.
% CRediT mapping of stated contributions: "YSZ grain growth data" ->
% Investigation, Data curation; "procedure" -> Methodology; "photos" ->
% Visualization; "editing" -> Writing - review & editing; O. Johnson (last
% author / PI) -> Supervision, Project administration.
\textbf{Ryan Weber:} Investigation, Methodology, Data curation, Visualization.
\textbf{Christopher Nyborg:} Investigation, Methodology, Data curation,
Visualization.
\textbf{Ronald Guymon:} Visualization, Writing -- review \& editing.
\textbf{Gage Erickson:} Visualization, Writing -- review \& editing.
\textbf{Oliver Johnson:} Supervision, Project administration.

\section*{Data Availability}

The data that support the findings of this study are openly available in the
development repository
(\url{https://github.com/vertical-cloud-lab/custom-induction-furnace}); an
archival snapshot is deposited at Zenodo,
\href{https://doi.org/10.5281/zenodo.20878017}{https://doi.org/10.5281/zenodo.20878017}.

\bibliography{references}
\end{document}